\documentclass[fp,twocolumn]{jpsj3}
\usepackage{txfonts}
\usepackage{bm}
\usepackage{graphicx}% Include figure files
\usepackage{color}

\title{A new inelastic neutron spectrometer HODACA}

\author{Hodaka Kikuchi$^1$, Shinichiro Asai$^1$, Taku J. Sato$^2$, Taro Nakajima$^1$, Leland Harriger$^3$, Igor Zaliznyak$^4$, and Takatsugu Masuda$^{1,5,6}$\thanks{masuda@issp.u-tokyo.ac.jp}}
\inst{$^1$Institute for Solid State Physics, The University of Tokyo, Chiba 277-8581, Japan \\
$^2$IMRAM, Tohoku University, Sendai, Miyagi 980-8577, Japan \\
$^3$NIST Center for Neutron Research, National Institute of Standards and Technology, Gaithersburg, Maryland 20899 \\
$^4$Condensed Matter Physics and Materials Science Department, Brookhaven National Laboratory, Upton, NY 11973, USA \\
$^5$ Institute of Materials Structure Science, High Energy Accelerator Research Organization, Ibaraki 305-0801, Japan \\
$^6$ Trans-scale Quantum Science Institute, The University of Tokyo, Tokyo 113-0033, Japan }

\abst{
A new multiplex-type inelastic neutron scattering spectrometer, HOrizontally Defocusing Analyzer Concurrent data Acquisition spectrometer (HODACA), was recently developed and built at the C1-1 cold neutron beam port in JRR-3. 
The spectrometer is suitable for dynamics measurements in the energy range of $-1$ meV $\lesssim \hbar \omega \lesssim$ 7 meV, catering to a broad array of research fields in physics and material science. 
HODACA combines 24 detectors and 132 pieces of analyzer crystals and has an estimated measurement efficiency that is 70 times greater than the existing conventional triple-axis spectrometer at the C1-1 beam port.
The concept, design, specification, and results of commissioning experiments are described. 
}

\kword{Inelastic neutron scattering spectrometer, HODACA}
\begin{document}
\maketitle

\section{Introduction}
Neutron scattering is an indispensable experimental technique in various fields, including physics, chemistry, materials science, and engineering. In 1949, Nobel laureate C. G. Shull demonstrated the utility of neutron scattering for magnetic structure analysis by observing magnetic Bragg peaks in the antiferromagnet MnO~\cite{Shull1949}, using the spin of neutrons. Since then, in the field of liquids and soft matter, large cross section of hydrogen or deuterium, has greatly advanced structural analysis using the small angle neutron scattering technique~\cite{Stuhrmann74}. In engineering, neutron diffraction and radiography are utilized for non-destructive testing of engines, concrete, and other materials~\cite{Schillinger04,Zhang17}. In condensed matter physics, neutrons with wavelengths of the order of atomic spacing are employed to study collective excitations in crystals~\cite{Squires72} and magnetic materials~\cite{Collins69}. For example, in the presence of magnetic order, excitations are described as spin waves, which propagate as small fluctuations of spins. The dispersion relation, which relates the wavevector ${\bm q}$ and energy $E$ of spin waves, can be measured by neutron inelastic scattering (INS) spectrometers~\cite{GLsq}, revealing the dynamics of magnetic materials and determining the spin Hamiltonian. 
Since the proposal of spin liquids~\cite{qsl}, the dynamics of frustrated magnetic 
materials such as triangular~\cite{Facheris22}, Kagome~\cite{Han2012492}, and pyrochlore~\cite{Kimura13} lattices have been actively investigated. 
When the ordering of spin 
dipoles is hindered, nontrivial physical quantities such as vector chirality, scalar chirality, 
and spin nematicity can be expected to order. 
In some cases, small perturbations such as DM interactions~\cite{Elhajal02} and magnetic dipole interactions~\cite{Maksymenko15} play crucial roles in determining the ground state. 
In all these cases, the spin Hamiltonian of the system can be 
determined by dynamics measurements using INS spectrometers, 
and the investigation of nontrivial ground states is conducted.

The neutron triple-axis spectrometer (TAS) has been widely used in both inelastic and elastic scattering experiments since its development in the 1950s, establishing its position as a versatile and important spectrometer~\cite{tas}. 
In a traditional TAS, a neutron scattered by a sample crystal is analyzed by a single analyzer crystal and detected by a single detector. Accumulated knowledge and expertise over the years have enabled high signal-to-noise ratio measurements at specific points in wave vector - energy (${\bm q}$ - $E$) space. 
This is different for time-of-flight (TOF) chopper spectrometer, a type of advanced inelastic scattering spectrometers, which is mainly installed at pulse neutron sources such as ISIS, SNS, and J-PARC. This spectrometer analyzes neutron energy by measuring the flight time, allowing simultaneous measurements over a wide range of energies. 
By placing multiple detectors around the sample, it enables simultaneous measurements over a wide range of ${\bm q}$ space, which allows for efficient measurements in a broad $\bm{q}$ - $E$ space.
The multiplex-type TAS spectrometer is another type of the advanced spectrometers. 
This spectrometer increases the number of analyzers and detectors to enable efficient measurements and is often installed at constant neutron sources, such as FLATCONE~\cite{KEMPA2006} at ILL, MACS~\cite{Rodriguez_2008} at NIST, and CAMEA~\cite{GROITL201699} at PSI. 
Currently, the design, construction, and operation of multiplex-type TAS spectrometers employing various concepts are being carried out globally, yielding significant achievements in the wide range of fields including frustrated magnetism~\cite{Han2012492,Kimura13,Facheris22}, low-dimensional quantum magnets~\cite{Hong10,Umegaki15,Gitgeatpong17}, spin liquids~\cite{Nakatsuji12,Sears15}, molecular magnets~\cite{Mourigal14}, superconductivity~\cite{Chen19,Butch22}, heavy fermions~\cite{Song20}, Weyl semimetals~\cite{Gaudet21}, van der Waals materials~\cite{Trainer22}, and relaxors~\cite{Phelan14}.
However, in Japan, progress in multiplex-type TAS spectrometers has stagnated because the research reactor JRR-3, a constant neutron source, had been shut down since the Great East Japan Earthquake in March 2011 until February 2021. 
This has changed recently with the design and construction of an advanced multiplex-type TAS spectrometer for the above mentioned research fields at the C1-1 beam port of JRR-3. 
The Inverse Rowland Inelastic Spectrometer (IRIS) concept 
proposed by Harriger and Zaliznyak~\cite{HarrigerZaliznyak_NCNR2015,HarrigerZaliznyak_unpublished} has been selected. 
This concept was implemented in an instrument named HODACA, which stands for HOrizontally Defocusing Analyzer Concurrent data Acquisition. 
In this paper, design, simulation, and components of HODACA, as well as test measurements on standard samples are described. 

\begin{figure}
\begin{center}
\includegraphics[width=7cm,pagebox=cropbox,clip]{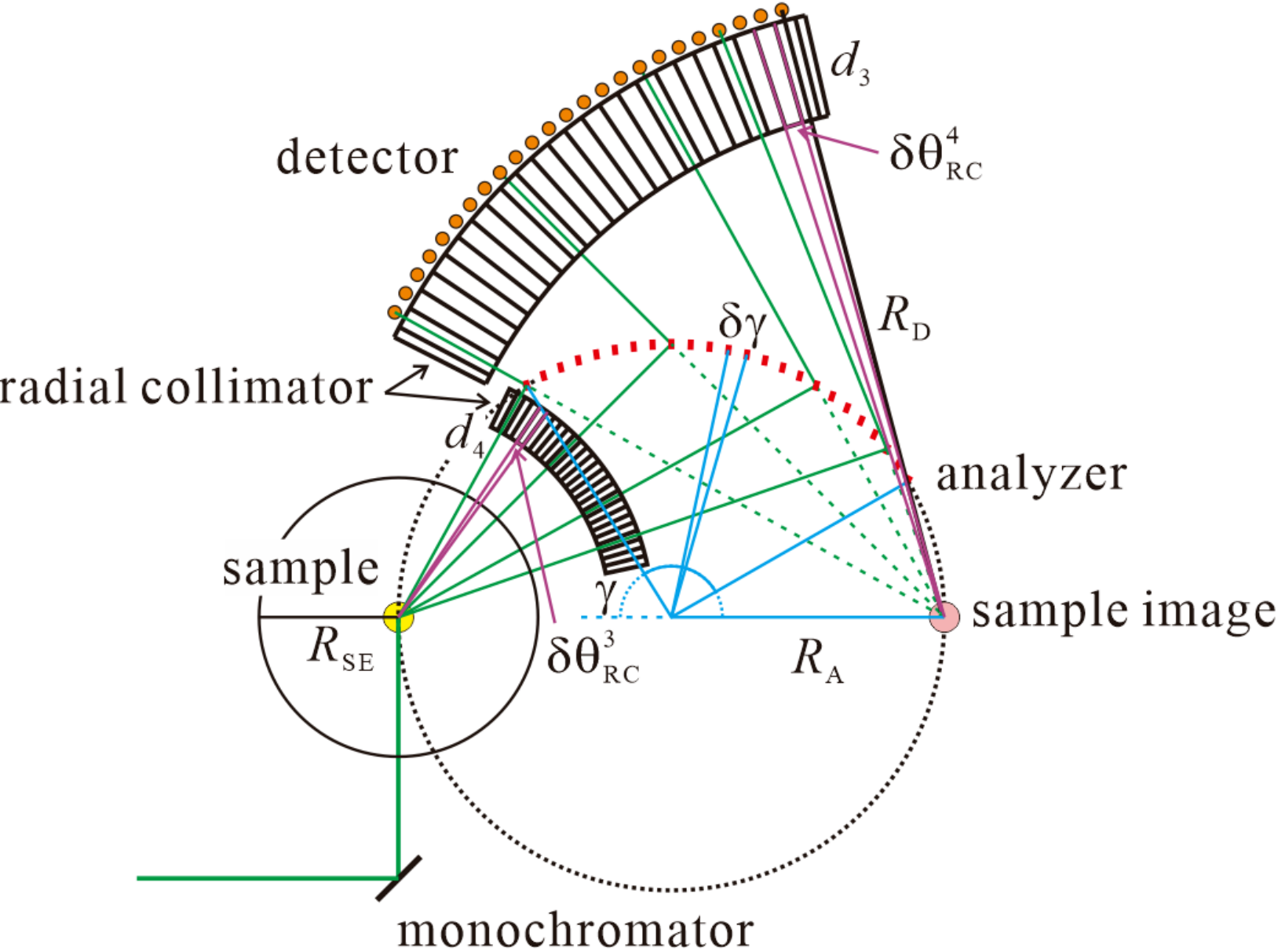}
\caption{Respresentative figure of HODACA spectrometer. Green lines indicate the neutron pathes.}
\label{f1}
\end{center}
\end{figure}

\section{Design of HODACA}
In HODACA, efficient collection of neutrons at the scattering plane is achieved through ``anti-focusing'' using an array of analyzers based on the concept of the Rowland circle~\cite{HarrigerZaliznyak_NCNR2015,HarrigerZaliznyak_unpublished} 
implemented as shown in Fig.~\ref{f1}. Neutrons scattered from the sample are reflected at a fixed angle, corresponding to specific scattered neutron energy, by the array of analyzers located on the Rowland circle. According to the circumferential angle theorem, the neutron path after reflection is depicted as spreading out 
from the sample image on the circumference (anti-focusing). The reflected neutrons are then detected by a group of detectors arranged around the sample image. To suppress cross-talk between neutrons before and after passing through the analyzer array, a radial collimator is placed before and after the analyzer group, which is expected to efficiently reduce background noise. 
Previous work using wide-angle analysis with position-sensitive detector has shown that usage of background-reducing radial collimator is absolutely essential for successful operation of such a setup \cite{ZaliznyakLee_MNSChapter,Zaliznyak_JAP2002}.
Based on this principle, HODACA enables high signal-to-noise ratio measurements in a wide ${\bm q}$ space at constant energy ($E$) surfaces.
Monte Carlo simulation software McStas~\cite{Lefmann199910,Willendrup202123} was used for the design of the neutron scattering spectrometer, where various parameters were tuned to calculate the spectra and compare the intensities, energy resolution, and angular resolution. Optimization was performed based on these comparisons. 
As a result, HODACA became a spectrometer capable of measuring spectra from $-1$ meV to 7 meV by fixing the scattered neutron energy $E_f$ at 3.635 meV. 
It consists of 24 analyzers and 24 detectors spaced at 2$^{\circ}$ intervals, covering a scattering angle A2 of 46$^{\circ}$. Each analyzer is composed of 3 to 7 PG crystals mounted in a vertically focusing configuration. 
The vertical size of each analyzer (number of PG crystals) is determined to ensure that the solid angles spanned by the analyzer viewed from the 
sample position are the same. 
Radial collimators with divergence angle of 2$^{\circ}$ are installed between the sample-analyzer and analyzer-detector to minimize cross-talk of scattered neutrons from neighboring analyzers. 
The optimized parameters of the analyzers setup are summarized in Table~\ref{t1}. 

\begin{table}[tb]
\centering
\caption{Parameters of the optimized HODACA. The definitions of the parameters are as shown in Fig.~1.}
\begin{tabular}{|c|c|} \hline
$R_{\rm SE}$ & 350 mm \\ \hline
$R_{\rm A}$ & 530 mm \\ \hline
$R_{\rm D}$ & 1220 mm \\ \hline
$\gamma$ & $58^{\circ}~\sim~154^{\circ}$ \\ \hline
$\delta\gamma$ & $4^{\circ}$ \\ \hline
$d_{3}$ & 100 mm \\ \hline
$\delta\theta^{3}_{\rm RC}$ & 2$^{\circ}$ \\ \hline
$d_{4}$ & 200 mm \\ \hline
$\delta\theta^{4}_{\rm RC}$ & 2$^{\circ}$ \\ \hline
\end{tabular}
\label{t1}
\end{table}

\section{Simulation of HODACA and HER}

\begin{figure}
\centering
\includegraphics[width=7cm,pagebox=cropbox,clip]{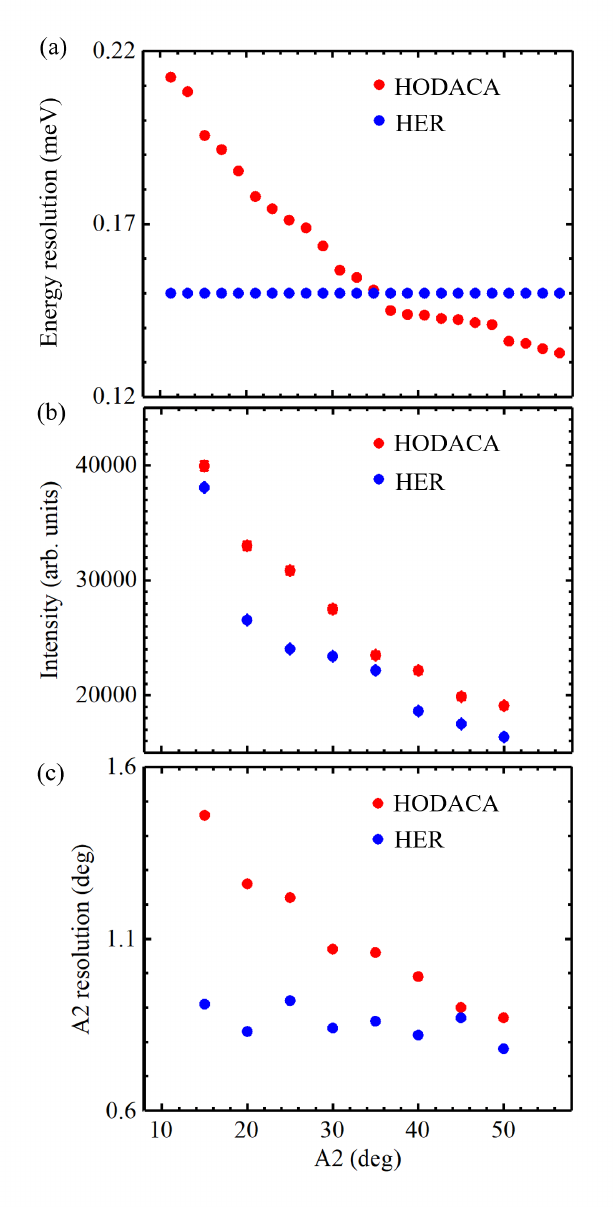}
\caption{Performance comparison between HODACA (red symbols) and HER (blue symbols). Plots show (a) energy resolution, (b) Bragg peak intensity for virtual a sample, and (c) angular resolution.}
\label{f2}
\end{figure}

The comparison of the performance between the optimized HODACA and HER is shown in Fig.~\ref{f2}. 
The simulation for HODACA was performed with a configuration covering $10^{\circ} < {\rm A2} < 56^{\circ}$. 
The collimator condition of HER was Ni guide - 200' - 80' - 80', and the analyzer mode was flat. 
First, let's compare the energy resolution as shown in Fig.~\ref{f2}(a). HODACA exhibits a location-dependent energy resolution ranging from 0.13 to 0.21 meV. 
This is due to the varying distances between the sample-analyzer and analyzer-detector, which are fixed at 0.8 m and 0.6 m, respectively. 
In contrast in HER, the energy resolution is constant at 0.15 meV, since the sample-analyzer distance and analyzer-detector distance remain constant. 
The maximum energy resolution of 0.21 meV in HODACA is equivalent to the energy resolution achieved in HER when the analyzer-detector collimator is 200'. Therefore, it is shown that at least one channel of analyzer-detector in HODACA has a performance that is superior to or equal to the collimator condition of Ni Guide - 200' - 80' - 200' in HER, in terms of energy resolution.

Figures~\ref{f2}(b) and \ref{f2}(c) represent the intensity of Bragg peaks and the angular resolution, respectively, for virtual powder sample. 
Here, the term ``virtual sample'' refers to a sample that produces the same magnitude of the structure factor for all Bragg reflections scattered at angles from 15$^{\circ}$ to 50$^{\circ}$ in 5$^{\circ}$ increments.
The Bragg peak intensity simulated for HODACA is higher than that for HER. This is due to the wider angular resolution of HODACA, ranging from 0.8$^{\circ}$ to 1.6$^{\circ}$, as shown in Figure~\ref{f2}(c), compared to the angular resolution of 0.8$^{\circ}$ in HER. The angle-dependent distances between the sample-analyzer and analyzer-detector result in an angle-dependent angular resolution. The IRIS-type spectrometer is designed primarily for inelastic scattering experiments, and it is empirically known that angular resolution within 2$^{\circ}$ allows for analyzable measurement data in inelastic scattering experiments using cold neutrons. Therefore, the detector dependence of the angular resolution (Fig.~\ref{f2}(c)) is not a concern when conducting inelastic scattering experiments. It is shown that at least one pair of analyzer-detector in HODACA has a performance that is superior to or equal to the collimator condition of Ni guide - 200' - 80' - 80', in terms of angular resolution.
Simulation using McStas compared TAS and HODACA, indicating that a single pair of analyzer-detector in HODACA has similar intensity and resolution to a TAS, but with 24 analyzer-detector channels HODACA provides 24 times higher measurement efficiency.

\section{Components of HODACA}

\begin{figure}
\centering
\includegraphics[width=7.5cm,pagebox=cropbox,clip]{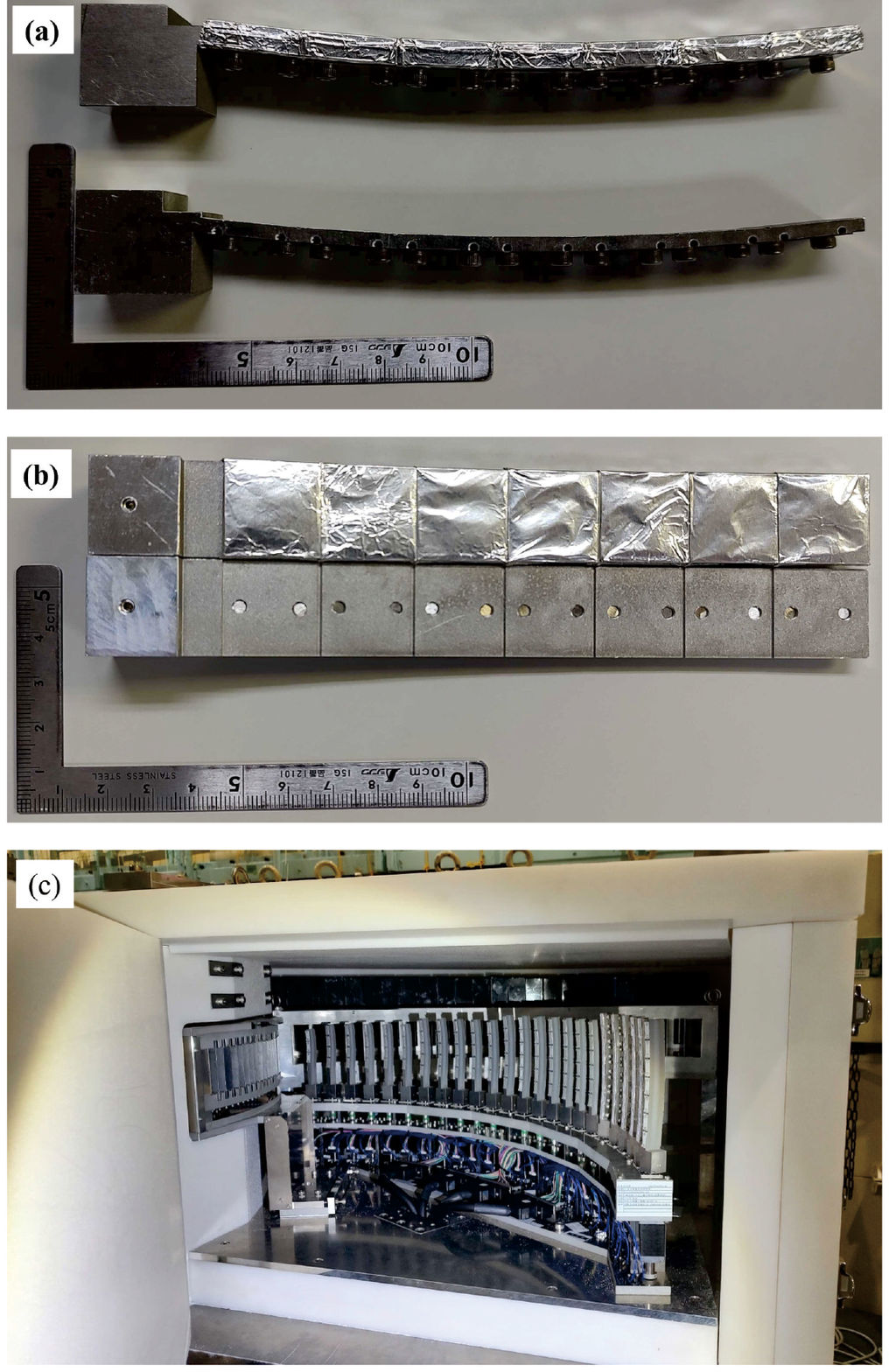}
\caption{ (a) A photograph of the analyzer viewed from the side. (b) A photograph of the analyzer viewed from the front. (c) A photograph showing the opened door of the shielding.}
\label{f3}
\end{figure}

Based on the design shown in Fig.~\ref{f1}, each component of the HODACA spectrometer was designed. 
(002) reflection of PG crystal is used for the analyzer. 
A total of 132 pieces, each measuring 20 mm $\times$ 20 mm $\times$ 2 mm, 
with a mosaic spread of 0.4$^{\circ}$ to 0.5$^{\circ}$, are used.
% (on account of secondary extinction this corresponds to an effective mosaic of $\sim$0.8 degrees for 3.7 meV neutrons). 
The HER analyzer is equipped with 49 PG crystals, which have the same mosaic spread as those used in HODACA. 
The crystals are arranged in a 7 by 7 configuration.
%The HER analyzer also uses an equivalent set of 49 pieces (7 columns $\times$ 7 rows). 
The PG crystals for HODACA are fixed to an aluminum shaft using aluminum foils (Figs.~\ref{f3}(a) and \ref{f3}(b)). 
Additionally, to suppress scattering by the aluminum shaft, Gd sheets are inserted between the PG crystals and aluminum shaft. 
To drive the 24 analyzers consisting of PG crystals and aluminum shaft, Oriental Motor CRK525PMAP motors and CRD5103P drivers are employed, for each of the 24 channels. The drivers are installed in the lower rack of the HODACA spectrometer. Two units of Tsuji Electronics Co., Ltd.'s 16-channel pulse motor controller (PM16C-16) are used to control these components.

For the detector bank, we used the spare detectors of AGNES spectrometer~\cite{KAJITANI1995872} in JRR-3. 
This detector is a $^3$He detector manufactured by Reuter Stokes, 
with a size of 1 inch diameter $\times$ 300 mm length, and operates as a 0-dimensional 
detector with 10 atmospheres of $^3$He gas pressure.

\begin{figure}
\centering
\includegraphics[width=8cm,pagebox=cropbox,clip]{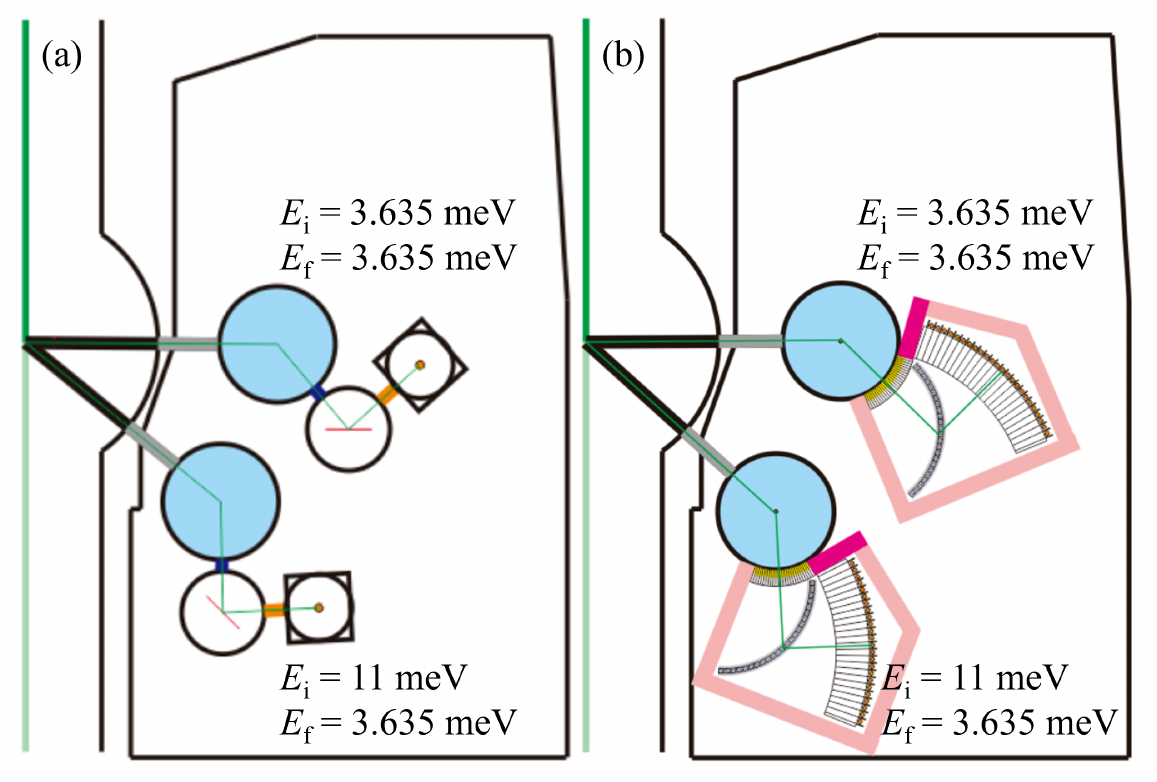}
\caption{(a) The top view (to scale) of the HER installed at the C1-1 beam port for $E_i$ = 3.635 meV and $E_i$ = 11 meV. (a) The top view (to scale) of the HODACA installed at the C1-1 beam port for $E_i$ = 3.635 meV and $E_i$ = 11 meV.}
\label{f4}
\end{figure}

\begin{figure}
\centering
\includegraphics[width=8cm,pagebox=cropbox,clip]{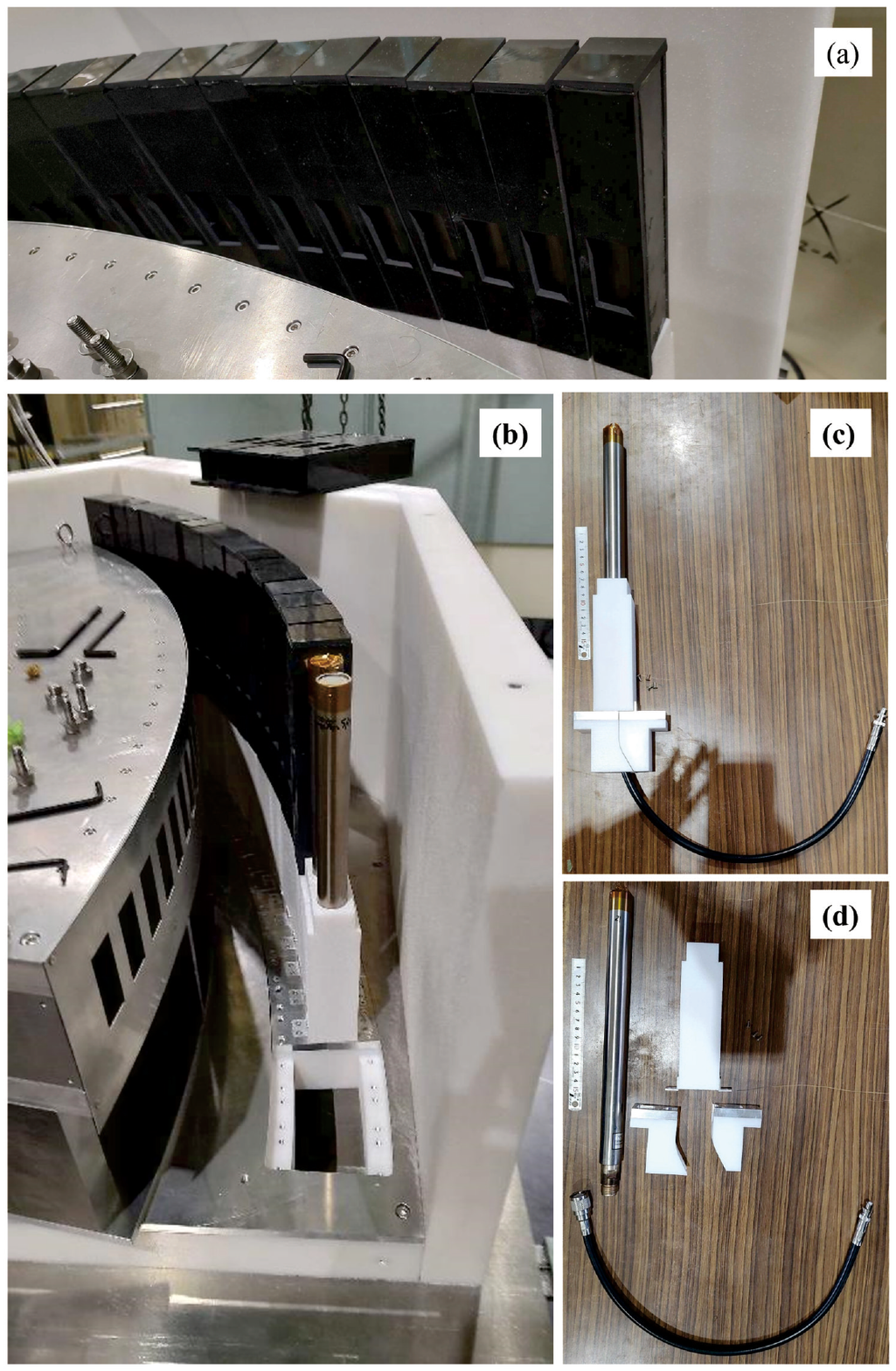}
\caption{(a) This is the detector cover made of B4C rubber. It has a window measuring 16 mm horizontally and 30 mm vertically. Additionally, at the rear of the fourth radial collimator, a Cd plate is attached, which also has a window measuring 16 mm horizontally and 30 mm vertically. (b) A photograph of the surroundings of the detector. (c) and (d) Photographs showing the detector configuration. In (c), the detector is assembled, while in (d), it is shown disassembled.}
\label{f5}
\end{figure}

The fabrication of the analyzer, detector, radial collimator, and shielding components was entrusted to Hitachi Denki Kogyo Co., Ltd. 
Figure~\ref{f4} is a scaled top view of the HER and HODACA installed at the C1-1 beam port. It shows the configurations for $E_i = 3.635$ meV and $E_i = 11$ meV. It can be observed that the measurement range covered by HER is also covered by HODACA. The shielding components of HODACA are indicated by vermilion color. Boron-10wt\% polyethylene is used for the shielding material. The darker vermilion areas indicate the possibility of direct neutron hits from the monochromator, for which boron-20wt\% polyethylene is used. The ceiling plate of the shielding components is designed to be easily removable to facilitate adjustments of the analyzer and detector. Additionally, the shielding components feature a door mechanism that allows immediate access for adjusting the analyzer bank (Fig.~\ref{f3}(c)). 
The detector is covered with a 4 mm thick boron-50wt\% B4C rubber and boron-10wt\% polyethylene (Fig.~\ref{f5}(b)). The B4C rubber has a 16 mm $\times$ 60 mm window, and a cadmium (Cd) plate is attached to the rear of the fourth radial collimator, also with a window size of 16 mm $\times$ 60 mm (Fig.~\ref{f5}(a)). These measures ensure that only scattered neutrons within the designed flight path are detected. The detector is securely fixed to the same boron-containing polyethylene pedestal as the shielding components, and the holes for wiring are curved to prevent neutron infiltration (Fig.~\ref{f5}(d)). The blades of the radial collimator are made of 1 mm thick Cd plates. Cd plates are very soft and prone to bending under gravity, so mechanisms are installed to attach each Cd plate with upper and lower ceiling plates. A crane is used to switch between HER and HODACA. The total weight of HODACA is 732 kg. HODACA is mounted on an aluminum frame, which includes the integrated radial collimator, analyzer, detector, and shielding components. To facilitate optical axis adjustment, the entire integrated assembly can be horizontally shifted.

The driving software used in HODACA is based on the SPectrometer and Instrument Control Environment, known as SPICE~\cite{LUMSDEN20061336}, which was modified for this purpose by CUSEY Co., Ltd. SPICE is commonly used for TAS control and allows intuitive and user-friendly control through simple commands. The main differences between HODACA and TAS are that in HODACA, control of C3 and A3 and control of vertical focusing are not required, and the output data includes counts for all 24 C3 and A3 channels. Here, C3 and A3 mean the rotation angle of an analyzer and the scattering angle of the analyzer, respectively.
Data display and analysis are currently being performed by handcrafted Python and Matlab codes. The development of dedicated software is underway.
% Dedicated software for HODACA data display and analysis, such as HANA~\cite{Kawana20181021} for HRC spectrometer in J-PARC or DAVE~\cite{Azuah2009114} for DCS in NIST has not been developed yet, and it is under construction. 

For noise reduction and signal amplification of the detector, two units of Clear Pulse 3341 16-channel counters and three units of 596 8-channel charge amplifiers are used. 
Currently, most of the aluminum components inside HODACA are exposed. However, plans are underway to add B4C rubber and Gd shielding to cover these components in 2023. Therefore, at present, it is expected that the background due to scattering from the aluminum components is significant. However, measurements using standard samples were conducted at this stage to establish the alignment direction, estimate the basic performance, and consider data display and analysis software. 

\section{Measurements of Standard Samples}
In this section, measurements of standard samples are described. 
The definitions of the motor angles and the details of the optical alignment in prior to the measurements 
are explained in Appendixes~\ref{appendix1} and \ref{appendix2}. 

\subsection{Vanadium}
We performed measurements using a cylindrical vanadium sample with a radius of 1 cm and a height of 4 cm. This measurement was carried out to determine the energy resolution and correct for the detection efficiency of the detectors.

\begin{figure}
\centering
\includegraphics[width=7cm,pagebox=cropbox,clip]{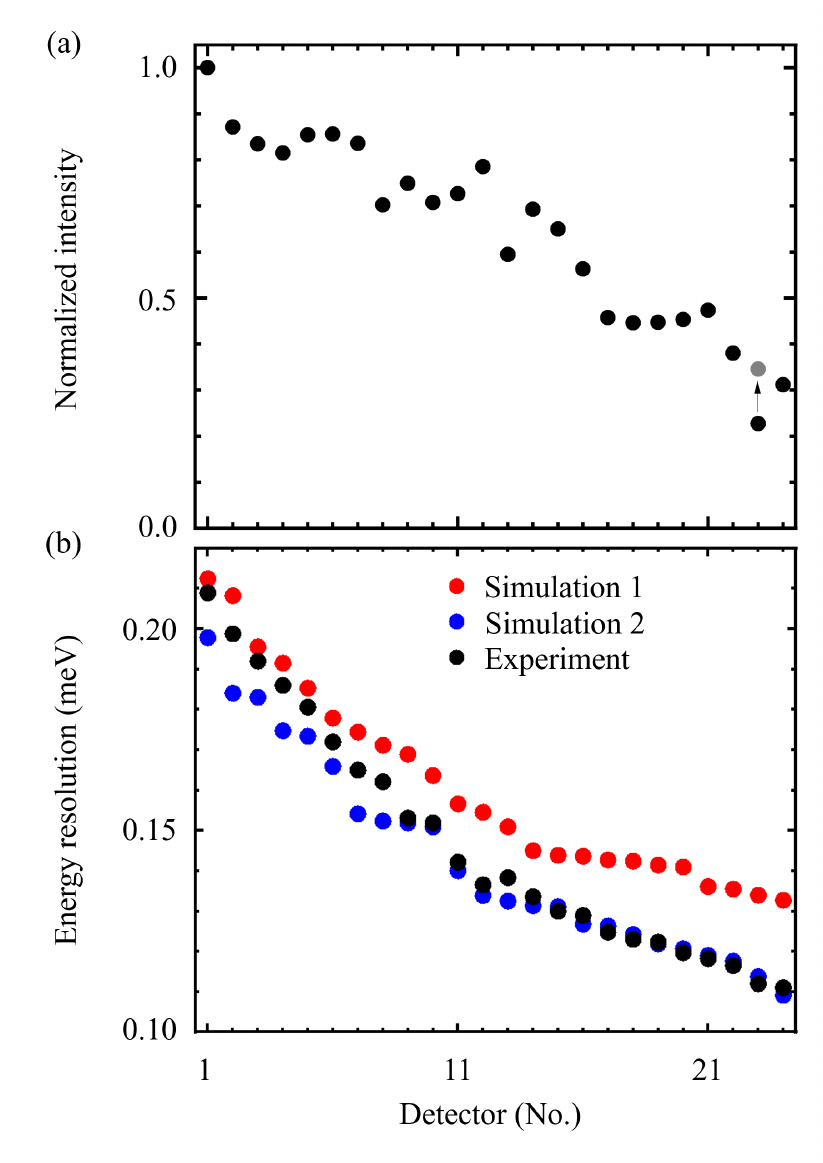}
\caption{(a) Intensity of vanadium measured by HODACA. The maximum intensity is normalized to 1. 
%\textcolor{red}{The black circles represent the measurements using vanadium as a reference. The gray dot correspond to the corrected intensity obtained from the measurements of Al$_2$O$_3$.} 
(b) Energy resolution estimated from Gaussian fitting of the full width at half maximum (FWHM). 
The black symbols represent the experimental results, while the red and blue symbols represent the simulation results for the monochromator and analyzer mosaic of 50' and 75', respectively. 
%The black symbols indicate the experimental values, while the red symbols represent the simulation results.
}
\label{f6}
\end{figure}

We conducted an energy scan in increments of 0.05 meV from $-0.2$ meV to 0.2 meV. The errors of the energy center were within 0.03 meV for all analyzers, which is within the motor tolerance range of A1. Here, A1 is the scattering angle of the monochromator as shown in Fig.~\ref{fa1} in appendix. We fit the data using Gaussian functions and estimated the intensity as shown 
in Fig.~\ref{f6}(a). 
The intensity is normalized with the maximum intensity set to 1. 
The detectors are numbered, starting with the one installed at the lowest scattering angle, which is number 1, and the numbers increase with the scattering angle. It can be observed that the scattering intensity of vanadium decreases as the detector number increases. 
There are two main reasons for this. First, the distance between the sample and the analyzer varies depending on the position of the analyzer, but the width of the analyzer crystals is fixed at 2 cm. As a result, the apparent angle of the analyzer width from the sample's perspective is not constant. Second, as the analyzer and detector numbers increase, both the sample-analyzer distance and the analyzer-detector distance become longer, leading to reduced scattering intensity due to air scattering.

Vanadium exhibits isotropic scattering, and, ideally, the same intensity is detected by all detectors if the flight distance of neutrons from the vanadium to the detector are the same. 
The measured scattering intensities obtained from each detector are calibrated by using the data in Fig.~\ref{f6}(a). 
The data point at detector number 23 was found to be incorrect during the measurements of Al$_2$O$_3$ in the forthcoming subsection. 
It was corrected to the gray circle using the scaling factor of Bragg profile of Al$_2$O$_3$.

In Fig.~\ref{f6}(b), the energy resolutions estimated from the full width at half maximum (FWHM) intensity in the experiment (black symbols) and simulation 1 (red symbols) are shown. 
For the simulation 1, the monochromator and analyzer mosaicities of 50', which was indicated by the specification of the commercial PG crystals, were used. 
Consistency is found in the detectors with smaller numbers. However, discrepancy grows with the increase in detector numbers. 
For the simulation 2, those of 75' were used. 
The consistency is better. 
The effective mosaicity of the monochromator and analyzer could be as large as 75' because of the secondary extinction of PG crystals. 

\subsection{Al$_2$O$_3$}

We conducted a diffraction experiment using Al$_2$O$_3$ as a standard sample with a radius of 10 mm and a height of 25 mm. 
This measurement allowed us to determine the value of A2$_n$ and to ascertain the presence or absence of cross-talk.
Here A2$_n$ is the scattering angle for $n$th analyzer and detector. The definitions of A2$_n$ as well as C1, A1, C2, C3n, and A3n are described in Appendix~\ref{appendix1}. 
%If cross-talk exists, false peaks should appear at angles shifted by 2 degrees. 

\begin{figure}
\centering
\includegraphics[width=8.5cm,pagebox=cropbox,clip]{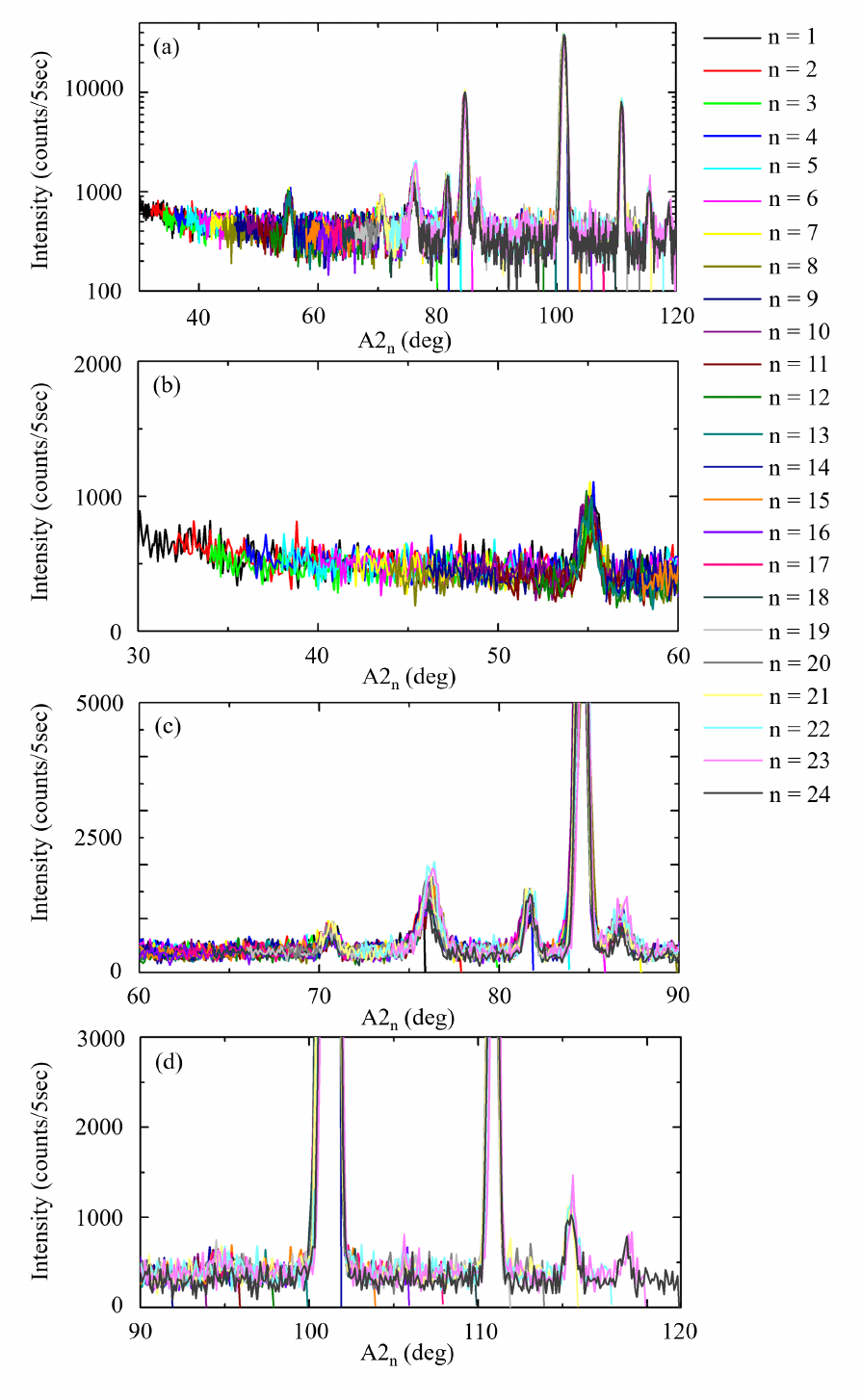}
\caption{The measured diffraction pattern of Al$_2$O$_3$. The intensity is corrected using Fig. 6(a), and the offset of each analyzer (A2) is also corrected. The diffraction pattern is presented with a logarithmic scale on the vertical axis for the entire measured range in (a). The patterns are further magnified in the ranges of 30$^{\circ}$ to 60$^{\circ}$ in (b), 60$^{\circ}$ to 90$^{\circ}$ in (c), and 90$^{\circ}$ to 120$^{\circ}$ in (d).}
\label{f7}
\end{figure}

At $E_i$ = 3.635 meV, a fewer number of Bragg peaks were detected. Therefore, we performed the measurement using the second-order harmonics of neutrons, which are neutrons with $E_i$ = 15.54 meV. This was made by removing the Be/PG filter between the monochromator and the sample. We measured neutron counts at A2$_n$ with 0.1-degree intervals with a duration of 5 seconds per point. The results are shown in Fig.~\ref{f7}, where the intensity has been corrected by vanadium data shown in Fig.~\ref{f6}(a). Additionally, we have also corrected offsets of A2$_n$, where the offsets are presumed to be the same for all A2$_n$. As shown in Fig.~\ref{f7}, the intensities and peak positions from all A2$_n$ match, meaning that the design of the spectrometer is sound. If the detector had detected reflections from adjacent analyzers, false peaks would have appeared at angles shifted by 2 degrees from the true angles. No such false peaks were observed in Fig.~\ref{f7}, confirming the absence of cross-talk. 

\subsection{Frustrated Magnet CsFeCl$_3$}

We have selected a frustrated magnet CsFeCl$_{3}$ as the standard sample for the measurement of 
an INS spectrum. 
The compound is an easy-plane-type antiferromagnet with one-dimensional chains of effective $S$ = 1 spins forming a triangular lattice. 
It belongs to the P6$_{3}$/mmc space group with lattice parameters of $a$ = $b$ = 7.2355 ${\rm \AA}$, $c$ = 6.0508 ${\rm \AA}$, $\alpha$ = $\beta$ = $90^{\circ}$, and $\gamma$ = $120^{\circ}$. 
Orbital moment of Fe$^{2+}$ ion is not quenched, and the ground state of the compound is known to be a quantum 
disordered state. 
Even though the magnitude of the magnetic moment has not been reported for CsFeCl$_{3}$ at ambient pressure, 
the reported magnitude of 3.15 $\mu_{\rm B}$ in the pressure-induced 120$^{\circ}$ structure at 2.2 
GPa~\cite{Hayashida18} would be a good reference value. 
The observed dispersion relations by INS at ambient pressure was adequately explained by the Extended Spin Wave Theory (ESWT)~\cite{Hayashida20195,Stoppel2021104}. 
The spin Hamiltonian is as follows:
\begin{eqnarray}
\mathcal{H}&=&J_{1}\sum_{\rm{n.n.}}^{\rm{chain}} {\bm S_{\rm i}} \cdot {\bm S_{\rm j}}~+~J_{2}\sum_{\rm {n.n.n.}}^{\rm{chain}} {\bm S_{\rm i}} \cdot {\bm S_{\rm j}}~+~~J_{3}\sum_{\rm{n.n.}}^{\rm{plane}} {\bm S_{\rm i}} \cdot {\bm S_{\rm j}} \nonumber
\\ 
&+&D\sum_{\rm i}^{site} (S^{z})^{2}, 
\label{eq:CsFeCl$_{3}$_hami}
\end{eqnarray}
where $J_1$ is the nearest-neighbor interaction along the c-axis, $J_2$ is the next-nearest-neighbor interaction, 
$J_3$ is the the nearest-neighbor interaction within the $ab$ plane, 
and $D$ is the easy-plane anisotropy.
The estimated parameters are $J_{1}=-0.271(1)~{\rm meV}$, 
$J_{2}=0.045(2)~{\rm meV}$, 
$J_{3}=0.016(1)~{\rm meV}$, and 
$D=2.144(4)~{\rm meV}$~\cite{Stoppel2021104}. 

The single crystal sample of CsFeCl$_3$ was grown using the vertical Bridgman method~\cite{Kurita201694}. 
The samples, weighing 1.64 g and 0.84 g, were prepared.
The crystals were coaligned using the transmission Laue method with a high-energy X-ray camera, ensuring that the [001] and [110] axes were in the horizontal plane. 
The coaligned crystals were fixed in the standard aluminum cell of Institute for Solid State Physics, the university of Tokyo. 
Cd shield for bulk aluminum or stainless steel screws of the cell was not installed. 

\begin{figure}
\centering
\includegraphics[width=9cm,pagebox=cropbox,clip]{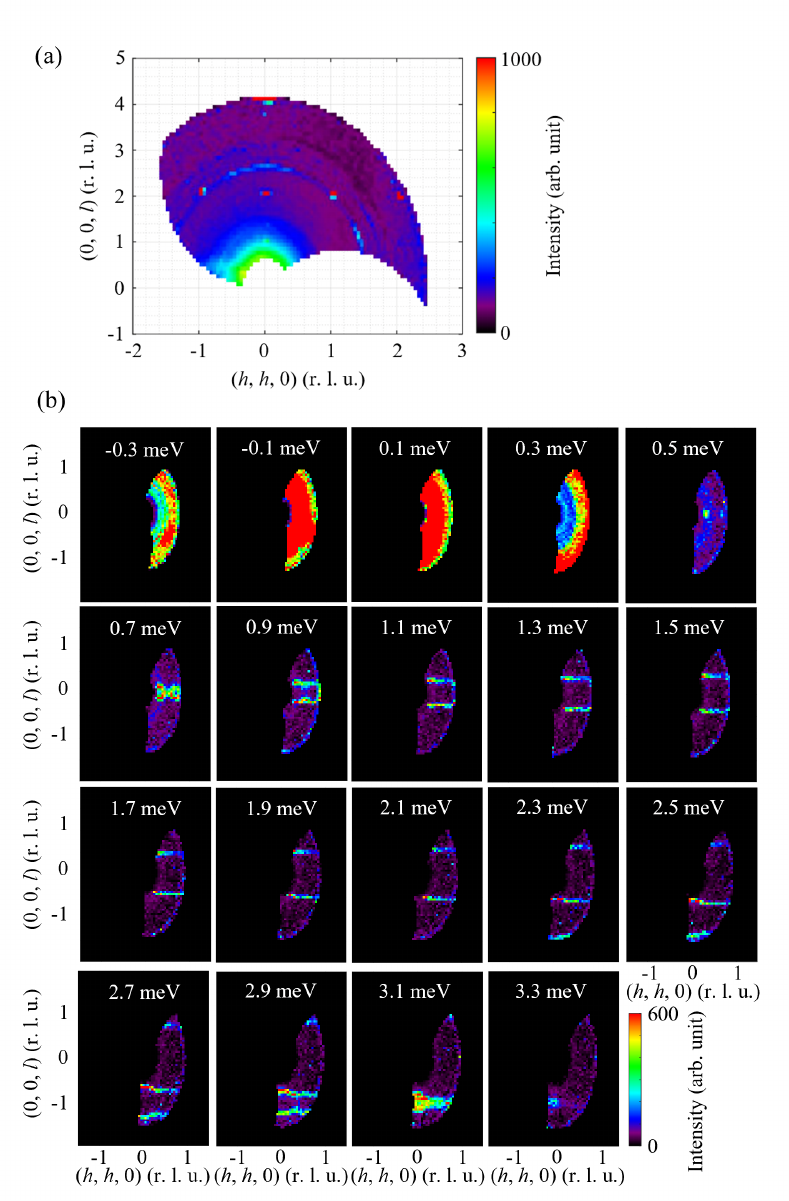}
\caption{(a) False color plot of the elastic scattering in CsFeCl$_3$ measured using the second-order harmonics of the neutrons with $E_i$ = 3.635 meV. (b) INS spectra in the range from -0.3 meV to 3.3 meV with increments of 0.2 meV.}
\label{f8}
\end{figure}

Neutron scattering measurements on CsFeCl$_3$ were performed at 2.4 K using GM refrigerator. 
The beam limiting aperture was installed in front of sample, and it was fully open. 
Firstly, diffraction profile was measured using the second order harmonics of the neutron with $E_i$ = 3.635 meV, corresponding to 14.54 meV. 
The coverage of the A2 angle ranged from 14$^{\circ}$ to 56$^{\circ}$ and from 58$^{\circ}$ to 104$^{\circ}$, while that of the C2 angle was within the range of 135$^{\circ}$. 
Figure~\ref{f8}(a) shows the results of the measurement. 
Bragg peaks were observed at six points: $(-1~-1~2)$, (0 0 2), (1 1 2), (2 2 2), and (0 0 4). 
According to the extinction rule, Bragg peaks were not observed for odd values of $l$. 
These results confirmed that the data reduction was accurate and there were no issues with the spectrometer setup or sample alignment.

\begin{figure}
\centering
\includegraphics[width=8cm,pagebox=cropbox,clip]{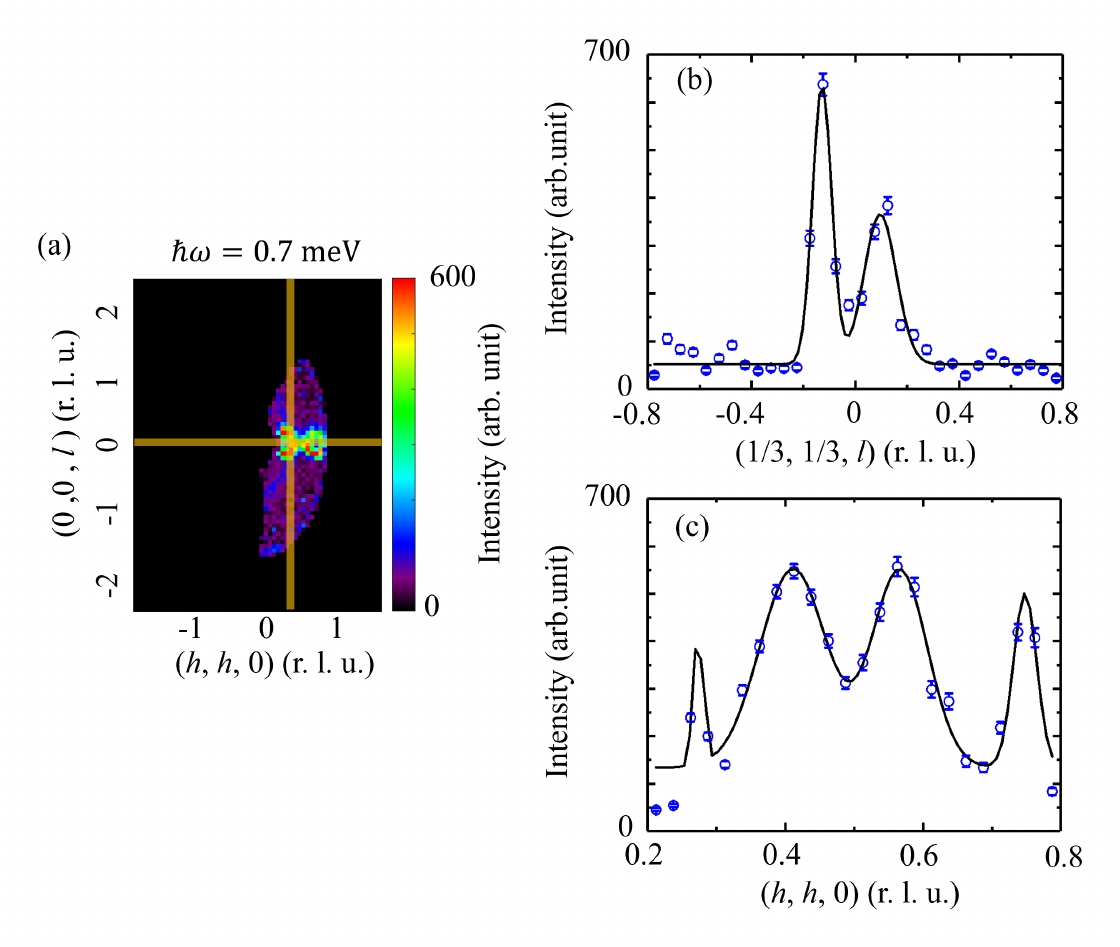}
\caption{(a) INS spectrum at $\hbar \omega$ = 0.7 meV. (b),(c) 1D cut along (b) $(1/3, 1/3, l)$ and (c) $(h, h, 0)$ direction at $\hbar \omega$ = 0.7 meV. The ranges $(1/3 \pm 0.05, 1/3 \pm 0.05, l)$ and $(h, h, 0 \pm 0.05)$ highlighted in yellow rectangles in (a) are integrated. The blue circles represent experimental data, and the black curves represent Gaussian fitting.}
\label{f9}
\end{figure}

Next, we measured the INS spectra. 
The measurements were conducted with a 0.05 meV interval in the range of $-0.4$ to 2.9 meV, and with a 0.1 meV interval in the range of 3.0 to 3.4 meV. 
The coverage of the A2 angle ranged from 14$^{\circ}$ to 60$^{\circ}$, while that of the C2 angle was within a range of 138$^{\circ}$ with a 2$^{\circ}$ interval. 
A total of 70 points (corresponding to the C2 steps) were measured per energy slice. 
Each point was measured for 1 minute, resulting in approximately 70 minutes to acquire the intensity map per energy slice. The total measurement time was approximately 84 hours. The data in the range from $-0.3$ meV to 3.3 meV with increments of 0.2 meV are shown in Fig~\ref{f8}(b). The binning size was set to 0.044 (r.l.u.) for both $(0, 0, l)$ and $(h, h, 0)$.

Figures~\ref{f9}(b) and ~\ref{f9}(c) represent one-dimensional (1D) cut along ${\bm q} = (1/3,~1/3,~l)$ and $(h,~h,~0)$ directions for the obtained energy slice at $\hbar \omega$ = 0.7 meV. 
The integration ranges are $(1/3 \pm 0.05, 1/3 \pm 0.05, l)$ for the former and $(h, h, 0 \pm 0.05)$ for the latter 
as indicated by yellow rectangles in Fig.~\ref{f9}(a). 
The data indicated by blue open circles in Figs.~\ref{f9}(b) and \ref{f9}(c) are reasonably well fitted by 
multiple Gaussians indicated by the black solid curves. 

\begin{figure}
\centering
\includegraphics[width=8cm,pagebox=cropbox,clip]{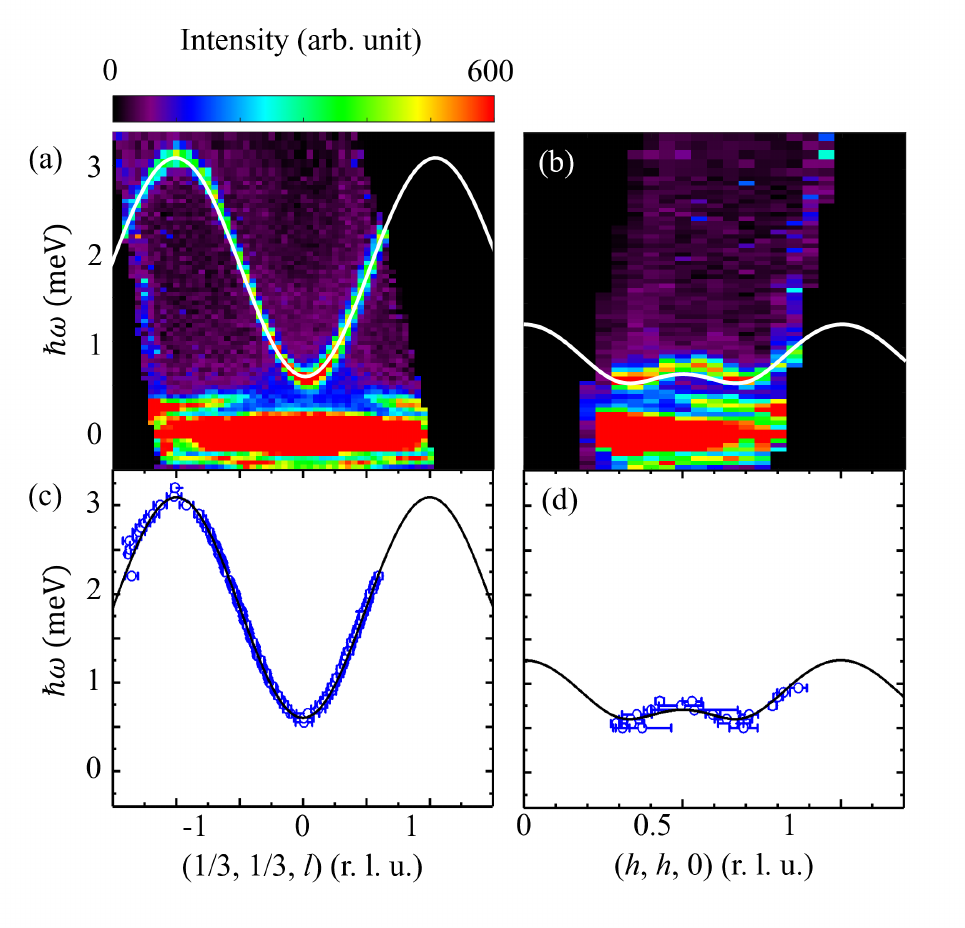}
\caption{(a) False color plots for the $(1/3, 1/3, l)$ direction and (b) $(h, h, 0)$ direction. The integration ranges are $(1/3 \pm 0.05, 1/3 \pm 0.05, l)$ and $(h, h, 0 \pm 0.05)$, respectively. The white dashed curves represent the dispersion curves obtained from ESWT.(c) and (d) show Gaussian fits of the constant-energy scans in the range of 0.5 to 3.4 meV for the $(1/3, 1/3, l)$ and $(h, h, 0)$ directions, respectively. The blue circles indicate the peak center positions, and the error bars in the momentum direction represent FWHM/2. The black curves represent the dispersion curves obtained from ESWT.}
\label{f10}
\end{figure}

Figures~\ref{f10}(a) and \ref{f10}(b) illustrate the false color plots for the INS spectra sliced by 
$\hbar \omega - {\bm q}$ plane, where ${\bm q} = (1/3, 1/3, l)$ for the former and $(h, h, 0)$ for the latter. The integration ranges are $(1/3 \pm 0.05, 1/3 \pm 0.05, l)$ and $(h, h, 0 \pm 0.05)$, respectively. The white curves in the figures represent the dispersion relation calculated using ESWT with the reported parameters~\cite{stoppelPRB}. 
Blue open circles in Figs.~\ref{f10}(c) and \ref{f10}(d) display the dispersion relations obtained by fitting the 1D-cuts along ${\bm q} = (1/3,~1/3,~l)$ and $(h,~h,~0)$, respectively. 
The error bars in the ${\bm q}$ directions indicate FWHM/2 from the fitting. 
The black curves represent the calculation obtained from ESWT. 
The experimental results are well reproduced by the calculations using the previously reported parameters. 

Strong background was observed in the low energy range of $\hbar \omega$ = $-$0.3 meV and 0.3 meV. 
Even though the origin was not clear at the time, the installation of Cd shield to the sample cell and adequate 
adjustment of the beam aperture reduced the background significantly in a recent experiment. 
\begin{figure}
\centering
\includegraphics[width=8cm,pagebox=cropbox,clip]{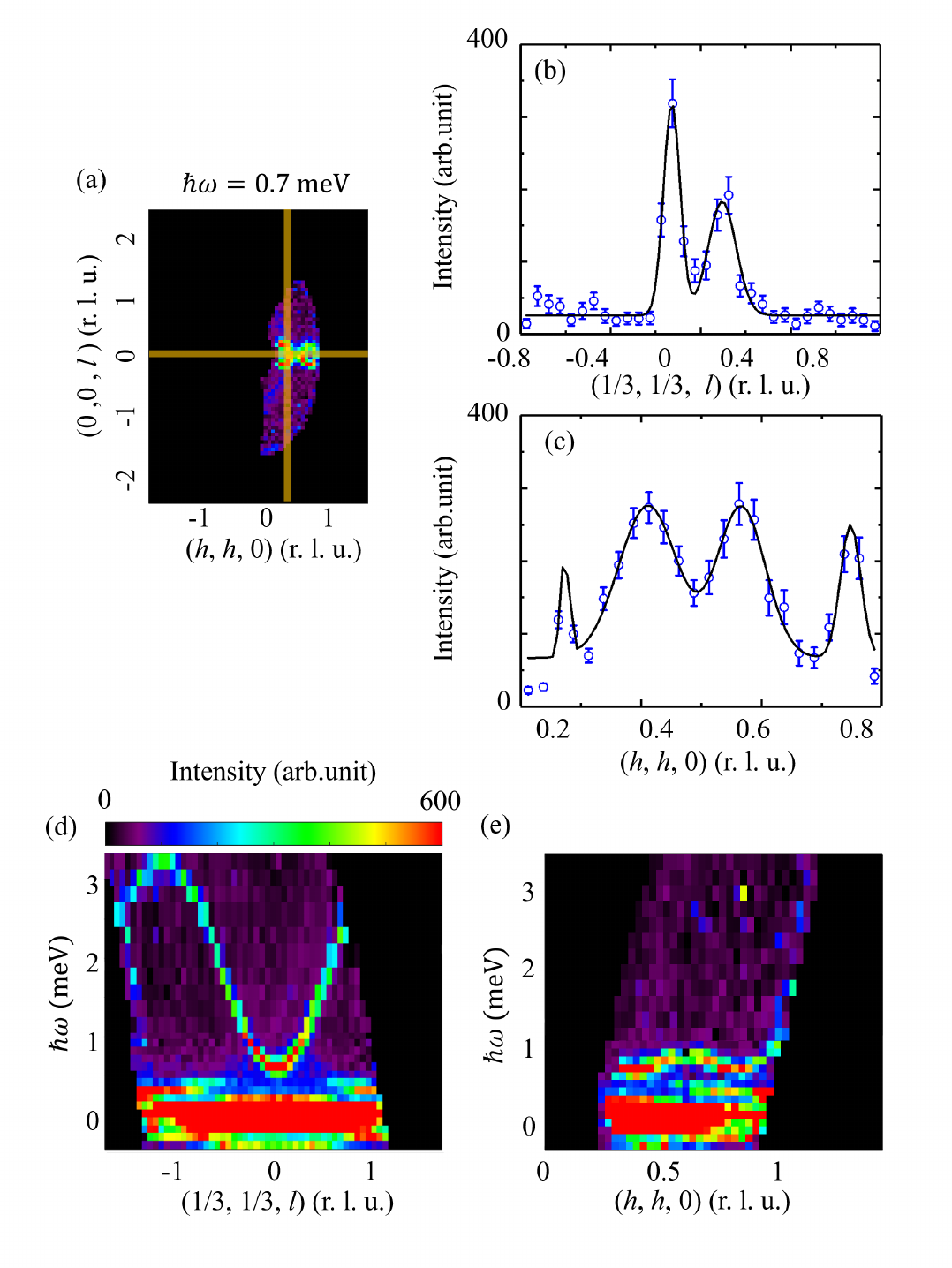}
\caption{(a)-(c) The spectrum (a) and 1D cuts (b),(c) were produced under the same analysis conditions as those in Fig. 9(a)-(c), assuming a counting time of 30 seconds for each point.
(d),(e) The spectra were produced under the same analysis conditions as those in Fig. 10(a) and (b), assuming the energy step sizes were modified to be 0.1 meV increments from -0.4 meV to 1.2 meV, and 0.2 meV increments from 1.2 meV to 3.4 meV.}
\label{f11}
\end{figure}

The obtained results were sufficiently analyzable even without scanning in the A2 direction. 
The measurement efficiency of HODACA on the energy slice surpasses the conventional TAS, such as HER, by a large margin. 
Here we will simulate how much we could reduce the measurement time. 
Let us reduce the measurement time per point to 30 seconds, half of the original experiment, and the results are shown in Figs.~\ref{f11}(a)-\ref{f11}(c). 
The error bars increase by a factor of $\sqrt{2}$. 
The fittings to the data in Figs \ref{f11}(b) and \ref{f11}(c) are still good. 
Figures \ref{f11}(d) and \ref{f11}(e) show the results obtained when the energy step size was changed to increments of 0.1 meV from $-0.4$ to 1.2 meV, and increments of 0.2 meV from 1.2 to 3.4 meV, respectively. 
Although the resolution is slightly coarse, sufficient data for determining the spin Hamiltonian can still be obtained. 
The data shown in Fig. \ref{f11} can be obtained in approximately 16 hours and 30 minutes by halving the measurement time and widening the energy step size. 
Even if all measurements were performed with the step size of 0.1 meV, the measurements would be completed in approximately 23 hours and 40 minutes, which means that the data of sufficient quality can be obtained within a day.
Typical beam time assigned to one proposal for an INS experiment in JRR-3 ranges from 5 to 7 days, and using HODACA enables efficient measurements.

\subsection{Recent measurement on Al$_2$O$_3$}
\begin{figure}
\begin{center}
\includegraphics[width=7.5cm,pagebox=cropbox,clip]{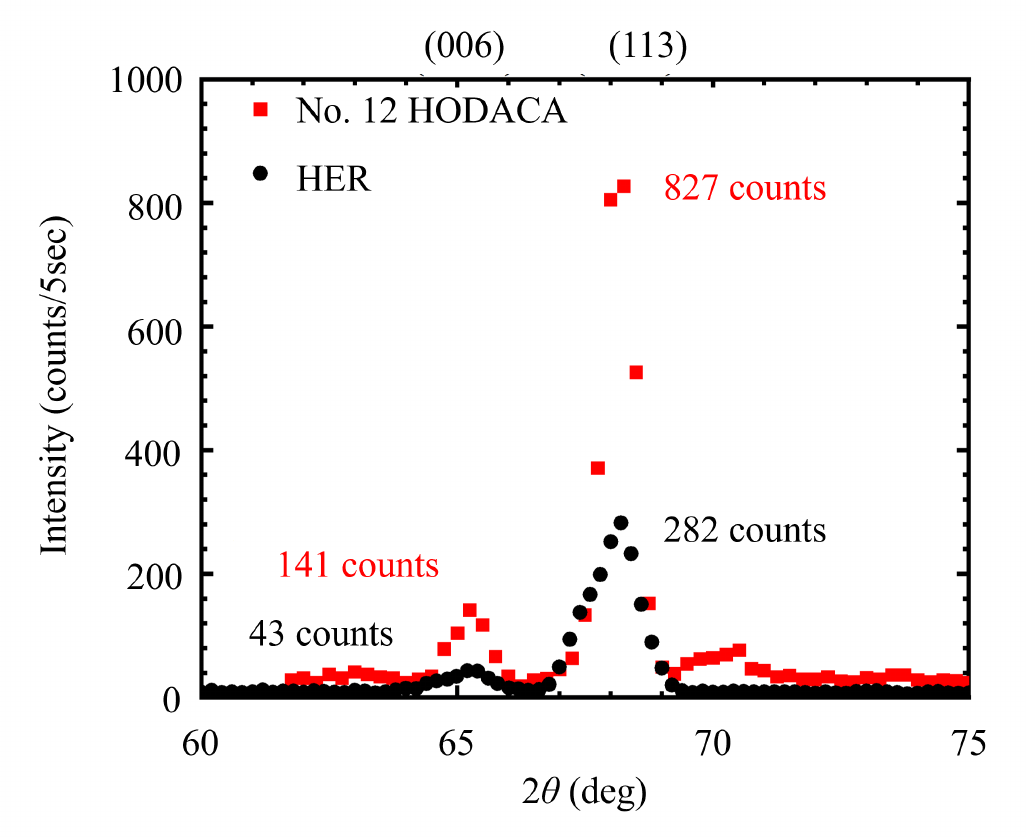}
\caption{Bragg peak profiles measured by HER (black symbols) and HODACA (red symbols). 
Collimator condition for HER was Ni guide - open - 80' - open. }
\label{f12}
\end{center}
\end{figure}

Recently, we reduced the background by adjusting the electronics. We then measured the Bragg peak profile of Al$_2$O$_3$ to compare the intensities of HODACA and HER. As demonstrated in Fig.~\ref{f12}, the intensity measured by the No. 12 detector in HODACA, represented by red symbols, is three times larger than that measured by HER, denoted by black symbols. The measurement of detailed data for the intensity comparison is currently under way, but early indications suggest that the efficiency of the HODACA detector might be approximately three times that of HER. Given that the intensity of the No. 12 detector is close to the average of the other detectors, as illustrated in Fig.~\ref{f6}, and that HODACA can be considered an assembly of 24 HER spectrometers, as discussed in section 3, we estimate that the measurement efficiency of HODACA could be up to 70 times higher than that of HER.

The background shown in Fig.~\ref{f12} is still large, though it has improved compared to that shown in Fig.~\ref{f7}. A supplementary radial collimator is currently being installed, with further improvements to the background expected.

\section{Summary}

In summary, we have designed and built a multiplex-type inelastic neutron scattering spectrometer HODACA at C1-1 beam port in JRR-3. 
The estimated measurement efficiency of HODACA is 70 times greater than that of an existing conventional triple-axis spectrometer at the beam port. 
An ideal spectrometer for measuring dynamics in the energy of cold neutrons is now ready for users. 

\begin{acknowledgment}

\acknowledgment
\section*{Acknowledgements}

We greatly appreciate Osamu Yamamuro for providing the spare $^3$He tube detectors of the AGNES spectrometer. Our heartfelt thanks also go out to Kenji Nakajima, Taka-hisa Arima, Tsuyoshi Kimura, and Yusuke Tokunaga for their valuable comments. We are also grateful to Makoto Ozeki for providing single crystals of CsFeCl$_3$. Special thanks to Ryosuke Sugiura, Toshio Asami and Daichi Kawana for their technical support in JRR-3.
Hodaka Kikuchi was supported by the Japan Society for the Promotion of Science through the Leading Graduate Schools (SpringGX). 
This project was supported by US-Japan cooperative program on neutron scattering. 
This project was additionally supported by JSPS KAKENHI Grant Numbers 19KK0069, 20K20896 and 21H04441. 
The work at Brookhaven was supported by the Office of Basic Energy Sciences, U.S. Department of Energy (DOE) under Contract No. DE-SC0012704.

\end{acknowledgment}

\appendix
\section{Definition of Angles in HODACA}\label{appendix1}

\begin{figure}
\centering
\includegraphics[width=7cm,pagebox=cropbox,clip]{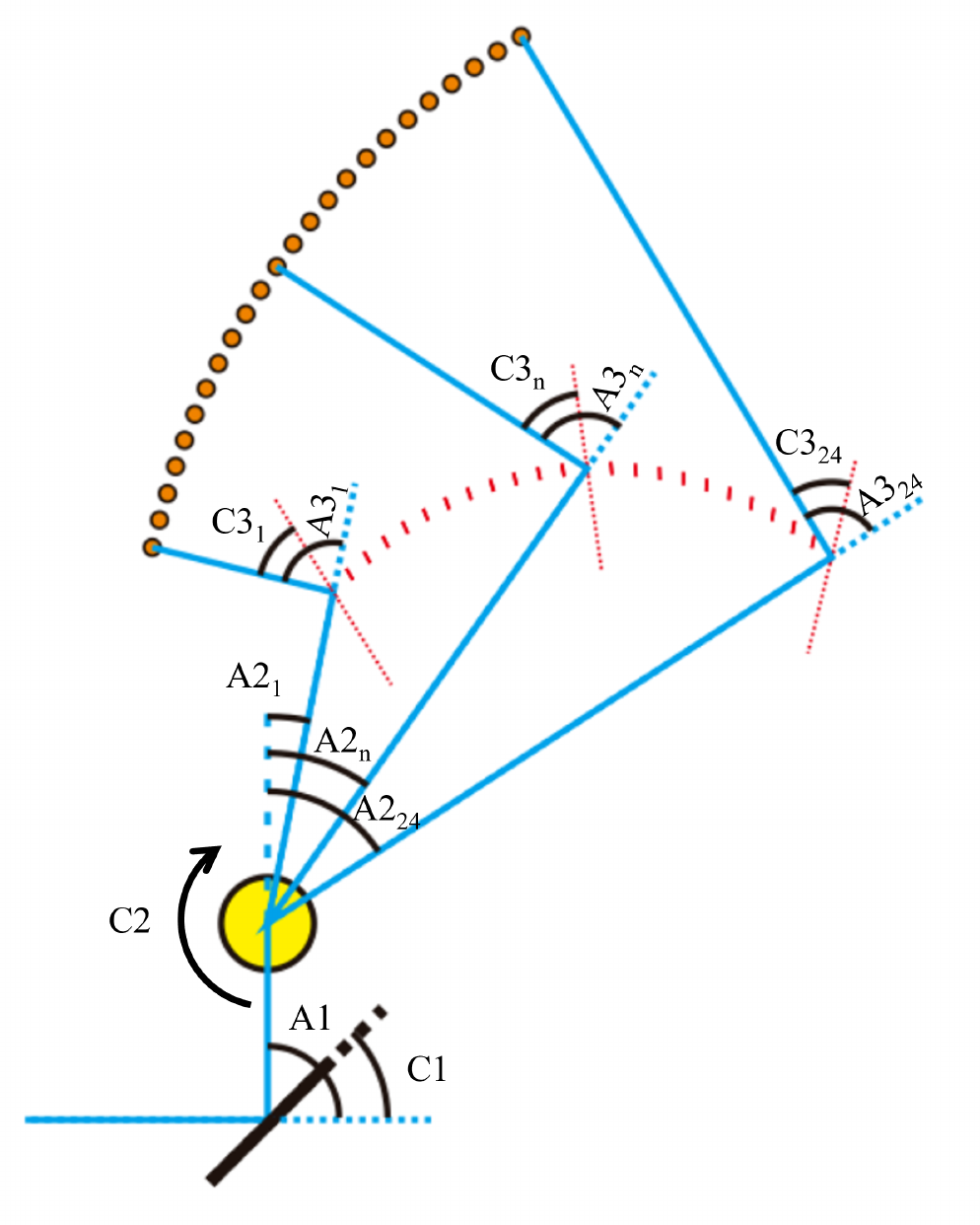}
\caption{Definition of angles in the HODACA spectrometer. The black line represents the monochromator, the yellow circle represents the sample goniometer, the red line represents the analyzer, and the orange circle represents the detector. The blue line indicates the flight path of neutrons.}
\label{fa1}
\end{figure}

The definition of angles in the HODACA spectrometer is depicted in Fig.~\ref{fa1}. Unlike a conventional TAS, HODACA has 24 rows of analyzer, resulting in the existence of 24 channels for sample scattering angle A2, analyzer rotation angle C3, and analyzer scattering angle A3. Therefore, the analyzers are numbered consecutively starting from the lower angle of the sample scattering, such as 1, 2, ..., 24, and the corresponding angles for the nth analyzers are defined as A2$_n$, C3$_n$, and A3$_n$, respectively. Similarly, the detectors corresponding to each analyzers are also numbered accordingly.

\section{Optical alignment of HODACA}\label{appendix2}

%\begin{figure}
%\includegraphics[width=8cm,pagebox=cropbox,clip]{figa2.pdf}
%\caption{Results of C3 scan for each analyzer. The red dots represent the measured data, and the black lines indicate Gaussian fitting.}
%\label{fa2}
%\end{figure}

Before performing measurements on the standard sample, the optical alignment of the HODACA spectrometer was carried out. Firstly, the height of the HODACA analyzer was adjusted to align its center position with the beam center. The HODACA spectrometer has an air pad with a screw-type height adjustment mechanism. The laser level tool was used to ensure that the tips of the two center pins (sharp-pointed rods) installed on the sample stage and the HODACA spectrometer are aligned in a straight line.

Next, the optical alignment of the third radial collimator was performed. This involved tying thread crosses on both the upstream and downstream apertures of the collimator and adjusting the height to ensure that they align in a straight line with the laser beam from the previous setup. Additionally, the distance between the center pin on the sample stage and the intersection of the collimator aperture was adjusted to be 35 cm by adjusting the distance between the sample stage and the HODACA spectrometer.

The fourth radial collimator was fixed to the HODACA spectrometer during fabrication, making optical alignment impossible. 
However, the height was verified using the same procedure as the third radial collimator.
Finally, the optical alignment of C3 was performed to ensure that the arrangement of the sample-analyzer-detector satisfies the scattering conditions for PG crystals. This was accomplished using a cylindrical acrylic with a radius of 10 mm and a height of 40 mm. By rotating C3, the intensity of the acrylic was observed at angles that satisfy the scattering conditions. The observed peaks were described by Gaussian functions. This process was performed for all 24 analyzers. The results were fitted by Gaussian functions, and the analyzers were fixed at the fitted center position for C3. Consequently, all analyzers were aligned to $E_f$ = 3.635 meV.

\section{Detailed profiles of Al$_2$O$_3$}\label{appendix3}
\begin{figure}
\centering
\includegraphics[width=7.5cm,pagebox=cropbox,clip]{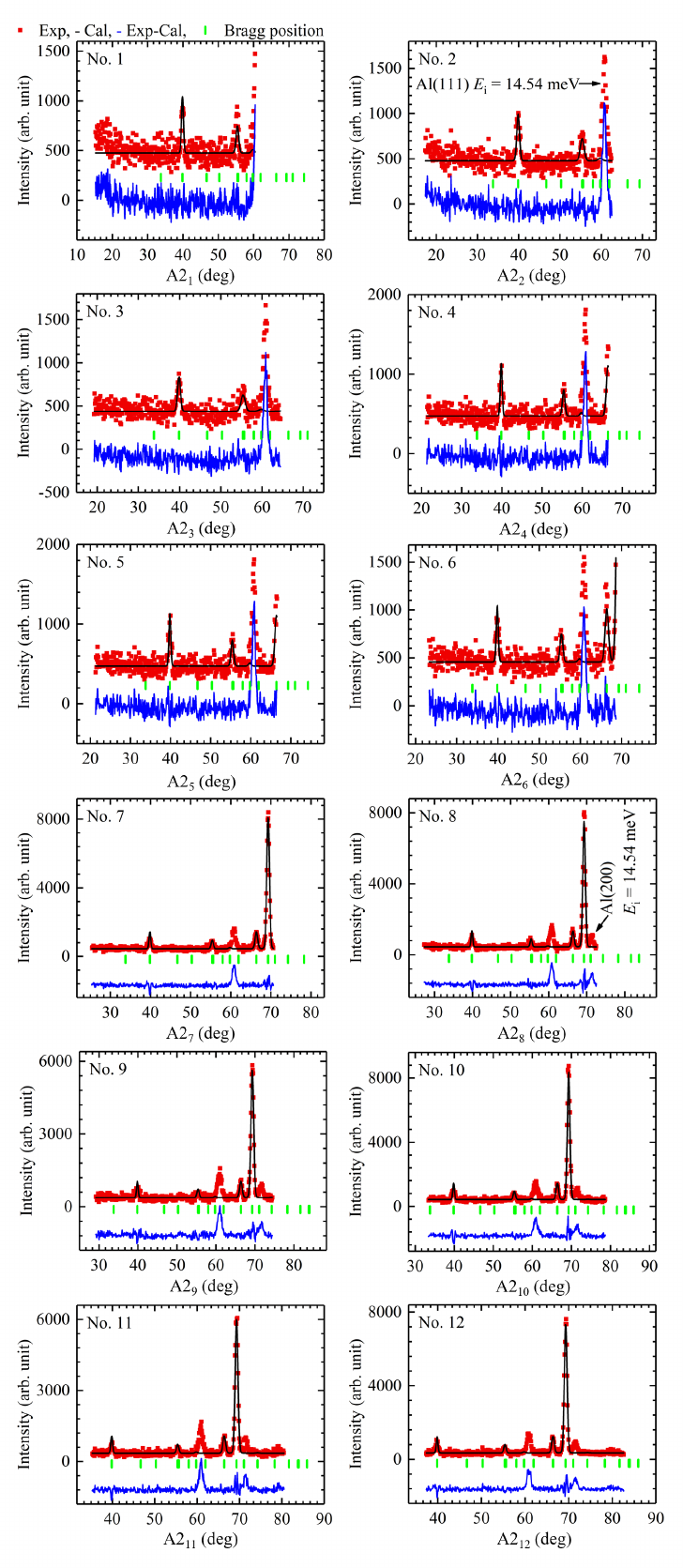}
\caption{Results of Rietveld analysis for each analyzer-detector pair. Shown are No. 1 to No. 12.}
\label{fa3}
\end{figure}

\begin{figure}
\centering
\includegraphics[width=7.5cm,pagebox=cropbox,clip]{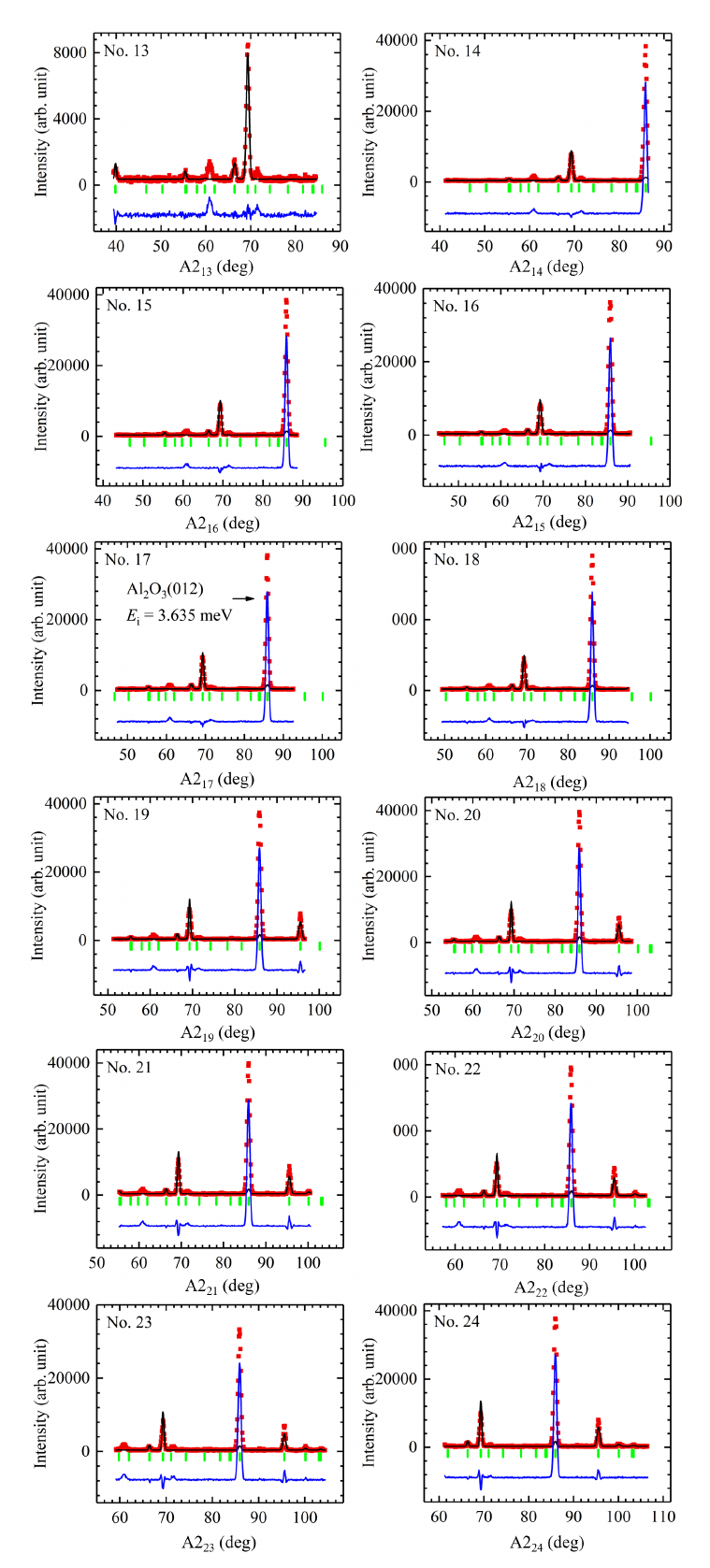}
\caption{Results of Rietveld analysis for each analyzer-detector pair. Shown are No. 13 to No. 24.}
\label{fa4}
\end{figure}

Figures~\ref{fa3} and \ref{fa4} show the results of the Rietveld analyses performed on the obtained 24 diffraction patterns using the Fullprof software~\cite{RODRIGUEZCARVAJAL199355}. 
%The analyses allow us to estimate the precise A2$_n$ offset for each analyzer. 
Most of the peaks were identified as reflections from Al$_2$O$_3$ with the neutron of $E_i$ = 14.54 meV, 
which is the second-order harmonics of the primary neutron beam of $E_i$ = 3.635 meV. 
Three peaks that did not match the experimental results were identified as reflections from Al(111) at 61 degrees with $E_i$ = 14.54 meV, Al(200) at 71 degrees with $E_i$ = 14.54 meV, and Al$_2$O$_3$(012) at 86 degrees with $E_i$ = 3.635 meV. 
An aluminum holder was employed for the Al$_2$O$_3$ sample, leading to the origin of the observed Bragg peaks of Al. 

%\textcolor{red}{
An assembly of analyzers and detectors is moved by a motor, A2$_0$, which is originally used for controlling the scattering angle of a sample in HER. 
The value of A2$_n$ when A2$_0$ = 0, which is defined as the offset for A2$_n$, was determined by the Rietveld analysis as shown in Fig. \ref{fa4}. 
The offset of a2$_n$ is linear with a slope of 2, which means that all analyzers were aligned as designed. 
Table \ref{ta1} summarizes the offset values and the deviations from their ideal positions. 
All the analyzers are positioned within 0.142 degree of the ideal locations. 
The reasons for this deviation is supposed to be the tilt caused by the Gd sheet and aluminum foil behind the PG crystal. 
This level of deviation does not cause cross-talk, and there are no issues with the arrangement of the analyzers.
%}

\begin{table}[tb]
\centering
\caption{Summary of the offset of A2$_n$ estimated from the measurement of Al$_2$O$_3$ and deviations from the ideal straight line.}
\begin{tabular}{|r|r|r||r|r|r|} \hline
No&A2$_n$ ($^{\circ}$)&Deviation ($^{\circ}$)&No&A2$_n$ ($^{\circ}$)&Deviation ($^{\circ}$) \\ \hline
1&-15.35&-0.088&13&8.79&0.057 \\ \hline
2&-13.22&0.038&14&10.70&-0.030 \\ \hline
3&-11.32&-0.061&15&12.74&0.013 \\ \hline
4&-9.30&-0.038&16&14.79&0.066 \\ \hline
5&-7.22&0.046&17&16.79&0.065 \\ \hline
6&-5.21&0.058&18&18.80&0.077 \\ \hline
7&-3.26&0.005&19&20.76&0.036 \\ \hline
8&-1.39&-0.129&20&22.74&0.020 \\ \hline
9&0.73&-0.009&21&24.73&0.010 \\ \hline
10&2.84&0.109&22&26.65&-0.072 \\ \hline
11&4.69&-0.046&23&28.58&-0.142 \\ \hline
12&6.78&0.044&24&30.69&-0.032 \\ \hline
\end{tabular}
\label{ta1}
\end{table}

\begin{figure}
\centering
\includegraphics[width=6cm,pagebox=cropbox,clip]{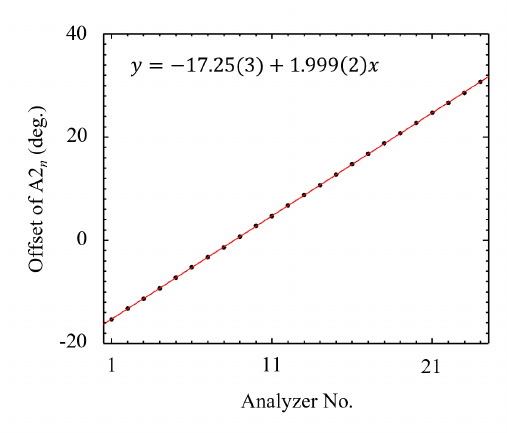}
\caption{
%\textcolor{red}{
The relationship between the offset of A2$_n$ and analyzer numbers. Black symbols represent measured values, and the red line indicates the fitted straight line.
%}
}
\label{fa5}
\end{figure}

%\bibliographystyle{jpsj} 
%\bibliography{refer}

%\begin{thebibliography}{9}
%\bibitem{jpsj} The abbreviation for JPSJ must be ``J. Phys. Soc. Jpn." \note{in the reference list}.
%\bibitem{instructions} More abbreviations of journal titles are listed in ``Instructions for Preparation of Manuscript".
%\bibitem{etal} The use of ``et al.'' is not accepted in principle, therefore, all the authors must be listed.
%\bibitem{ibid} The term ``ibid.'' should not be used even if the same journal or book is cited with different page numbers.
%\bibitem{Errata} Errata should be listed under the same reference number. 
%\end{thebibliography}

\end{document}